\documentclass[pra,twocolumn,showpacs,superscriptaddress]{revtex4-1}
\usepackage{amsmath,amscd,amssymb,color}
\usepackage{graphicx,amsfonts,epsf}
\usepackage{epstopdf}
\usepackage{float}
\usepackage{hyperref}
\usepackage{enumerate}
\usepackage{array}

\usepackage{mathtools}
\usepackage{breakurl}
\usepackage{xcolor}
\usepackage{ulem}
\begin{document}
\title{Experimental investigation  of coherence contributions to
a nonequilibrium thermodynamic process in a driven quantum
system}
\author{Krishna Shende}
\email{ph19032@iisermohali.ac.in}
\affiliation{Department of Physical Sciences, Indian
Institute of Science Education \& 
Research Mohali, Sector 81 SAS Nagar, 
Manauli PO 140306 Punjab India.}
\author{Kavita Dorai}
\email{kavita@iisermohali.ac.in}
\affiliation{Department of Physical Sciences, Indian
Institute of Science Education \& 
Research Mohali, Sector 81 SAS Nagar, 
Manauli PO 140306 Punjab India.}
\author{Arvind}
\email{arvind@iisermohali.ac.in}
\affiliation{Department of Physical Sciences, Indian
Institute of Science Education \& 
Research Mohali, Sector 81 SAS Nagar, 
Manauli PO 140306 Punjab India.}
\begin{abstract}
The work done when a system at thermal equilibrium is externally driven by a
unitary control parameter leads to 
irreversible entropy production.  The entropy produced can be thought
of as a combination of coherence generation and a population mismatch
between the target equilibrium state and the actually achieved final
state. We  experimentally explored this out-of-equilibrium process in
an NMR quantum processor and studied the contribution of coherence to
irreversible entropy generation. We verified a generalized Clausius
inequality, which affirms that irreversible entropy production is
lower-bounded.
\end{abstract} 
\maketitle 
\section{Introduction}
Non-equilibrium thermodynamics analyzes the behavior of systems that are far
from equilibrium, either because they are undergoing rapid changes or because
they are subject to external influences~\cite{Kawai_2007,Parrondo_2009}.  Such
non-equilibrium thermodynamics processes are usually accompanied by
irreversibility and entropy production, which has been
theoretically~\cite{Deffner_2010,Plastina_2014,Francica_2019,Santos2019} as
well as experimentally well studied on mechanical, NMR, and ion trap
setups~\cite{Koski2013,Batalhao_2014,Batalhao_2015,An2015,Smith_2018}.  Quantum
engines have been experimentally implemented in
NMR\cite{Peterson_2019,Assis_2019}, superconducting qubits\cite{PRX_IBM}, NV
centers~\cite{NV_engine_2019} and ion traps\cite{Trap1,Trap2}.  One of the key
steps in constructing these quantum engines consists of driving the system from
one equilibrium configuration to another, driven by an external 
Hamiltonian.
This process is ideal when the transformation is achieved
quasi-statically, but in reality this process is irreversible, which is
quantified using entropy.  The relative entropy produced in such a process can
be divided into two parts: the first part is the contribution coming from
coherence generation and other part can be attributed to the unwanted
transitions occurring after the unitary
transformation\cite{Francica_2019,Santos2019}.  The coherence produced is
quantified using the measures proposed in
Refs.~\cite{Baumgratz_2014,Streltsov_2017}.  The Clausius inequality, $\Delta
S_{irr} \ge 0$ states that the change in relative entropy is always greater
than zero.  In many cases, a process-specific bound is required, which can be
tighter than the Clausius inequality. This type of bound for the
non-equilibrium evolution of a quantum system is bounded by the Bures length
between two density operators\cite{kakutani_1948}.

In this work we study irreversibility using an NMR quantum processor, where an
NMR-active spin is initialized at thermal equilibrium and later subjected to a
unitary transformation to externally drive the system into a non-equilibrium state. The
driving time is changed from 100~$\mu$s to 800~$\mu$s, to probe the behavior of
entropy production with increasing driving time. We compute the irreversible
entropy produced during this process and compare it with the theoretically
expected values. We evolve the system at two different non-equilibrium states
by keeping its initial equilibrium temperature same. We separate the
contributions in entropy production coming from coherence generation and from
unwanted transitions.  We verify the inequality proposed by
Bures~\cite{Bures_length}, which states that entropy production is
lower-bounded by a non-zero positive value.

This paper is organized as follows: In Sec.\ref{sec2}, we 
briefly describe the theoretical
framework of irreversible entropy and 
the separation of contributions from
coherence and unwanted transitions. The Bures length inequality is also
defined. Sec.~\ref{sec3} describes the experimental implementation in detail.
Sec.~\ref{sec4} contains a discussion of the results
of the paper. A few concluding remarks are given
in Sec.~\ref{sec5}.
\section{Theoretical Background}
\label{sec2}
Recent research has focused on using coherence measures to
separate the contributions to irreversible work produced 
in a unitary quantum process  where the system is externally
driven into a nonequilibrium state.
The two contributions are those arising from coherence generation during the
process as well as from incoherent transitions.

Assume a closed system prepared in an initial thermal equilibrium state (at
inverse temperature $\beta_i$) via a Hamiltonian H[$\lambda(t)$] with a
controllable work parameter $\lambda(t)$.  The system is then subjected to a
Hamiltonian which changes from an initial $H_i$=$H[\lambda_i]$ to a final
$H_f$=$H[\lambda_f]$, in a finite driving time $\tau$.  The initial equilibrium
state is given by $\rho_i=e^{-\beta_i H_i}/Z[\lambda_i,\beta_i]$, where
$Z[\lambda_i,\beta_i]$ is the partition function, $\beta_i$ is the initial
inverse temperature and $\lambda_i$ is the work parameter. The time evolved
state is given by
$\rho_{\tau}=U_{\tau,0}[\lambda]\rho_iU^{\dagger}_{\tau,0}[\lambda]$, where
$U_{\tau,0} = T e^{-(i/\hbar)\int_{0}^{\tau}dtH[\lambda(t)]}$ and $T$ is the
time ordering operator.
Since no heat transfer takes place during this process, 
the irreversible work amounts to
the difference between the
average work ($\langle w \rangle$) 
and the free energy difference between the 
initial and final equilibrium 
states~\cite{Deffner_book}:
\begin{equation}
\langle w_{{\rm irr}} \rangle = \langle w \rangle -\Delta F
\label{W_irr}
\end{equation}
Further, Jarzynski's relation\cite{Jarzynski_1997}
can be used to derive $\langle w \rangle \ge \Delta F$.

\begin{figure}[h] 
\includegraphics[scale=1]{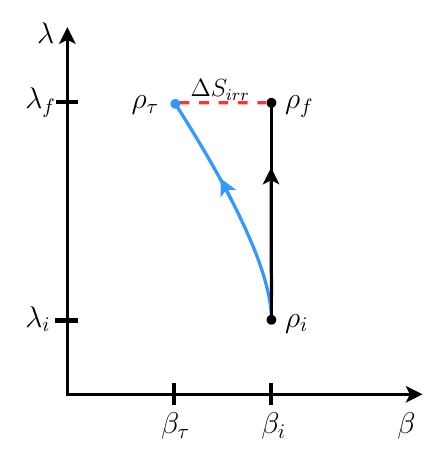} 
\caption{(Color online) Schematic diagram of a non-equilibrium quantum
thermodynamic process, with the system Hamiltonian H[$\lambda(t)$]
initially prepared in an equilibrium state $\rho_i$($\beta_i$,
$\lambda_i$) at $\tau$=0. The system is then pushed out of equilibrium
by changing the parameter $\lambda(t)$ to $\lambda_f$ during a time
$\tau$.  The irreversible entropy produced during this process is due
to the work difference between the actual state of the system
$\rho_{\tau}$ and the final equilibrium state $\rho_{f}$($\beta_i$,
$\lambda_f$). } 
\label{non-equilibrium_diag} 
\end{figure}

Consider a system initialized at $\rho_i(\beta_i,\lambda_i)$, which is driven
out of equilibrium for a time $\tau$. The actual state of the system is
$\rho_{\tau}(\beta_{\tau},\lambda_f)$ which is different from the ideal state
$\rho_f(\beta_i,\lambda_f)$, as shown in Figure~\ref{non-equilibrium_diag}.  
The irreversible
entropy during the transformation 
$\rho_i \rightarrow \rho_{\tau} $ 
is defined as:
~\cite{Deffner_2010}:
\begin{equation}
\Delta S_{irr} = \beta_i\langle w_{irr} \rangle = \beta_i(\langle w \rangle -\Delta F)=D(\rho_{\tau}||\rho_f)
\label{S_irr_dis}
\end{equation} 
where $D(\rho_{\tau}||\rho_f)$ is the quantum relative entropy\cite{Hisaharu}
between the actual density operator $\rho_{\tau}$ and the equilibrium density
operator.  The non-equilibrium lag in Eqn.~\ref{S_irr_dis} is a `proxy' for
entropy production, since, after the driving when the system is coupled to the
bath, it will relax from $\rho_{\tau}$ to
$\rho_f$~\cite{Spohn_1978,Breuer_2003,Santos2019}.
 
The irreversible entropy can be decomposed into two parts: (1) unwanted
transitions occurring between energy levels during the external driving which
makes the populations between $\rho_{\tau}$ and $\rho_f$ distinct, and (2) the
coherence generated in the actual state $\rho_{\tau}$~\cite{Francica_2019}: 
\begin{equation} 
\Delta S_{irr}  = C(\rho_{\tau}) + D(\Delta_{\tau}
[\rho_{\tau}||\rho_B])
\label{entropy} 
\end{equation} 
where the contribution due to coherence
generation is quantified using the relative entropy of coherence of
$\rho_{\tau}$ given by~\cite{Baumgratz_2014,Streltsov_2017}:
\begin{equation}
C(\rho_{\tau})=D(\rho_{\tau}||\Delta_{\tau}[\rho_{\tau}])=S(\Delta_{\tau}[\rho_{\tau}])-S(\rho_{\tau}).
\end{equation}
By construction, both the coherence and 
the mismatch in the population difference between actual and
final equilibrium states
are positive, hence the increase in either of 
these quantities leads to an
increase in total entropy production.

The Clausius inequality, $\Delta S_{irr} \ge 0$, provides a fundamental bound
on irreversible entropy production for a nonequilibrium thermodynamic process,
which is independent of how far from equilibrium the system reaches after the
process.  Using a geometric approach
for classical cases near equilibrium, the entropy produced is given by the
Riemannian distance between the initial and final states is given by
\cite{Salamon_1983,Salamon_1985,Nulton_1985,Ruppeiner_1995,Crooks_2007}:
\begin{equation} 
dS_{irr} \ge dl^{2}/2  
\end{equation} 
where the thermodynamic
length ($l$) has been shown to be the same as Wootters' statistical
distance between two pure states $\rho_{\tau}$ and $\rho_{f}$, given
by~\cite{Wootters_1981}: 
\begin{equation} 
l(p_{\tau},p_f)=arccos \left(
\int dx\sqrt{p_{\tau}(x)p_{f}(x)} \right) 
\end{equation} where $p_i$
and $p_{\tau}$ are the probability distributions of $\rho_0$ and
$\rho_{\tau}$. 
 
The Clausius inequality has been generalized to quantum nonequilibrium
processes by Deffner et.al~\cite{Deffner_2010}, which also places a lower bound
on the amount of entropy produced and is given by: 
\begin{equation} 
\Delta S_{irr}\ge
\dfrac{8}{\pi^2}L^2(\rho_{\tau},\rho_f) 
\label{general_inequality}
\end{equation} 
where $L$ is the finite Bures length, which is a quantum generalization of the
thermodynamic length and for two density operators, is given
by~\cite{Salamon_1983,Salamon_1985,Nulton_1985,Crooks_2007}:
\begin{equation}
L(\rho_1,\rho_2)=arccos\sqrt{F(\rho_1,\rho_2)} 
\end{equation} 
where $F$ is the
fidelity between two quantum states and is given 
by~\cite{Uhlmann_1976,Jozsa_1994}:
\begin{equation}
F(\rho_1,\rho_2)=\left[Tr\left(\sqrt{\sqrt{\rho_1}
\rho_2\sqrt{\rho_1} }\right)
\right]^2 
\end{equation} 
It is evident from 
Eqn.~\ref{general_inequality} that
quantum entropy production $\Delta S_{irr}$ is
bounded from below by the geometric distance between the actual density
operator and the equilibrium density operator. The further the
system is from equilibrium, the more is the entropy produced.
\section{Experimental implementation} 
\label{sec3} 
An NMR quantum processor consists of an ensemble of nuclear spins 
isotropically tumbling in a large static magnetic field, from which
qubits 
(spin-1/2 nuclei)
can be realized. In the high-temperature, high-field approximation,
NMR qubits follow a Boltzmann population distribution, which gives
rise
to a net magnetization along the $z$-direction (parallel
to the applied magnetic field). 
The NMR Hamiltonian for a two-qubit system in the rotating frame is
given by~\cite{OLIVEIRA}:
\begin{equation}
\mathcal{H}=-\hbar\sum\limits_{i=1}^2 { \omega _i } I_z^i + \hbar\sum\limits_{i<j=1}^2  J_{ij} I_z^i I_z^j
\end{equation}
where $\omega_i$ the offset frequency of the \textit{i}$^{th}$ nucleus, $I_z^i$
represents the $z$-component of the spin angular momentum of the
\textit{i}$^{th}$ nucleus and $J_{ij}$ is the scalar coupling between the
\textit{i}$^{th}$ and the \textit{j}$^{th}$ nuclei. 
 
\begin{figure}[h] 
\includegraphics[scale=1]{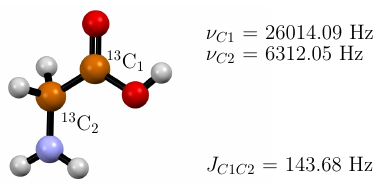} 
\caption{(Color online) Molecular structure of ${}^{13}$C-labeled glycine with
the $^{13}$C$_1$ and $^{13}$C$_2$ spins encoded as the first and second
qubit, respectively.  The offset rotation frequency for each spin and
scalar coupling strengths are listed alongside. } 
\label{mol-fig}
\end{figure} 

\begin{figure}[ht] 
\includegraphics[scale=1]{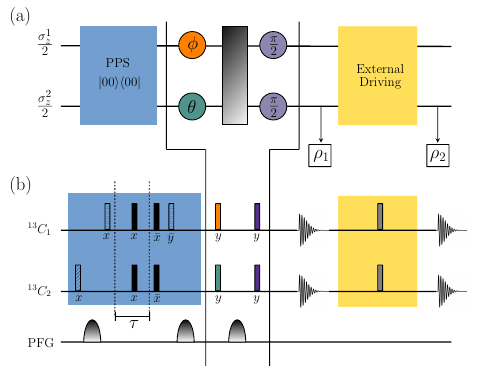} 
\caption{(Color online) (a) Circuit diagram to realize a non-equilibrium
quantum thermodynamic process on an NMR processor.  The  circuit to
achieve the pseudopure state (PPS) $|00 \rangle \langle 00|$ as shown
in the shaded blue box.  Orange and teal circles represent unitary
gates with angle of rotation fixed at $\phi$ and $\theta$ respectively,
which controls the equilibrium spin temperature of each qubit.  The
shaded yellow box depicts the external driving circuit which drives the
system out of equilibrium.  (b) NMR pulse sequence to implement
	the circuit given in panel (a).
Cross-hatched lined, horizontal lined and filled rectangles represent rf
pulses of $\frac{\pi}{3}$, $\frac{\pi}{4}$ and $\pi$ rotation angle
respectively, with their respective rotation axes given below each
shaded rectangle.  Orange, teal and indigo rectangles represent rf
pulses of $\phi$, $\theta$ and $\frac{\pi}{2}$ rotation angles
respectively, applied along the $y$-axis.  Grey shaded rectangles
correspond to external driving unitary pulses.  The PFG line depicts
the times at which a gradient is employed to destroy coherence. The
time delay $\tau$ is set equal to $\frac{1}{2J_{C_1 C_2}}$.  } 
\label{ckt_diag} 
\end{figure}

The molecule of 
${}^{13}$C-labeled glycine 
has been used to realize a system of two NMR qubits (Figure~\ref{mol-fig}).
The quantum circuit and corresponding pulse sequence to perform the
study of non-equilibrium dynamics is shown in Figure~\ref{ckt_diag}. 
The spatial averaging technique\cite{CORY1998} has been
used to  initialize the 
system in the pseudopure state $|00\rangle \langle 00|$, 
with the corresponding density operator
given by 
\begin{equation} 
\rho_{00}=\dfrac{1-\epsilon}{4}I_4+\epsilon|00\rangle
\langle 00| 
\end{equation} 
where $I_4$ is the 4 $\times$ 4 identity operator and $\epsilon$ is
proportional to the spin polarization ($\approx$ 10$^{-5}$ at room
temperature). The circuit and pulse sequence used to achieve this is shown in
the blue shaded portion of Figure~\ref{ckt_diag}.  Starting from this
pseudopure state, the qubits were initialized in the Gibbs thermal state with
the inverse temperature $\beta_i$ achieved by applying an rf pulse with an
angle of rotation between 0 and $\frac{\pi}{2}$, which redistributes the
population between the qubit states and also creates coherences.  To achieve
the final thermal state, wherein each spin is equilibrated at a different
pseudospin temperature, pulsed field gradient (PFG) pulses were applied to kill
the unwanted off-diagonal elements of the density matrix.  Quantum state
tomography was performed to to compute the state fidelity and to verify that
the system has been initialized to the required Gibbs state.  Gradient ascent
pulse engineering (GRAPE)~\cite{Khaneja_2005}, an optimal control technique,
was used to generate high-fidelity rf pulses of duration $\approx 150 \mu$~s to
implement single-qubit unitary rotations.  Pulses generated using GRAPE were
also used to drive the system out of equilibrium, which are shown in the yellow
shaded section of Figure~\ref{ckt_diag}.  All experiments  were  performed at
room  temperature on a Bruker Avance III 600-MHz FT-NMR spectrometer equipped
with a QXI probe. 
 
The effective spin temperature ($\beta_i$) of the initial Gibbs state is
related to the ground ($p_0$) and excited ($p_1$) populations by:
\begin{equation} 
\beta_i=\dfrac{1}{h\nu}\ln\left(\frac{p_0}{p_1}\right)
\end{equation} 
where $h$ is Planck's constant and $\nu$ is the offset frequency.  After
creation of the pseudopure state, the population exchange between the ground
and excited states can be controlled by applying an appropriate rf pulse,  o
initialize the system to the desired Gibbs state.  Both the spins were
initialized to an initial temperature of $(\beta_ih)^{-1}$=1580.2~Hz, where the
initial energy gap ($\nu$) was set to 2000~Hz.
 
The Hamiltonian which was used to drive the system out of equilibrium 
is given by~\cite{Batalhao_2015,Peterson_2019,Assis_2019}:
\begin{equation}
H(\nu(t))= -\frac{1}{2} h \nu(t)[\cos{(\pi t/2\tau)}\sigma_{x} 
+ \sin{(\pi
t/2\tau)}\sigma_{y}] 
\end{equation} 
where $\nu(t)=\nu_i(1-(t/\tau))+\nu_f(t/\tau)$. 
The energy gap can be expanded from $\nu_i$ at time $t=0$ to $\nu_f$ at
$t=\tau$ via implementation of the unitary($U_{\tau,0}$) 
calculated as~\cite{Assis_2019}:
\begin{equation}
U_{\tau,0}=exp\left(-i/\hbar\int_{0}^{\tau}H(\nu(t))dt\right).
\end{equation}
Non-equilibrium dynamics was probed by keeping the initial energy gap $\nu_i$
fixed at 2000~Hz and setting the $\nu_f$ to two different values, namely,
3600~Hz and 5000~Hz, respectively.  The driving time to drive the system into a
nonequilibrium state is varied from 100~$\mu$s to 800~$\mu$s.  This driving
time is much less than the decoherence times of the system qubits, and hence
the process can be considered to be almost unitary.

Quantum state tomography~\cite{Gaikwad2021,Gaikwad2022}
was performed
to confirm the closeness between the experimental and theoretical state and to
calculate irreversible work and coherence.  The 
density matrix corresponding
to an initial equilibrium state ($\rho_i$)  and to a final
non-equilibrium state ($\rho_f$) was 
reconstructed. The similarity between the theoretically
predicted density operator ($\rho_t$) and the experimentally reconstructed
density operator ($\rho_e$) is given  by the fidelity~\cite{Zhang_2014}:
\begin{equation}
F=\frac{|Tr(\rho_e
\rho_t^{\dagger})|}{\sqrt{Tr(\rho_e\rho_e^{\dagger})
Tr(\rho_t\rho_t^{\dagger})}}
\end{equation}
The high experimental fidelities
validate the good match between 
the theoretically expected and the experimentally
obtained values of irreversible work and coherence.
\section{Results and Discussion} 
\label{sec4} 
\begin{figure}[h]
\includegraphics[scale=1]{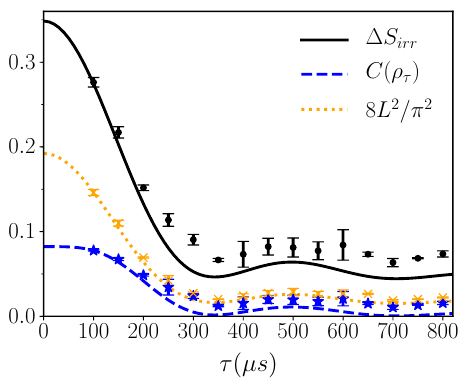} 
\caption{(Color online) Dynamics of entropy ($\Delta S_{irr}$), coherence
($C(\rho(\tau))$), and the lower bound on entropy (8L$^2$)/$\pi^2$,
plotted as a function of driving time($\tau$).  The energy gap is
varied from 2000~Hz to 3600~Hz.  Theoretical predictions are plotted as
solid curves, while experimental data (with error bars) are plotted as
circles, stars and crosses, respectively.
}	
\label{3600_plot} 
\end{figure}

\begin{figure}[h]
\includegraphics[scale=1]{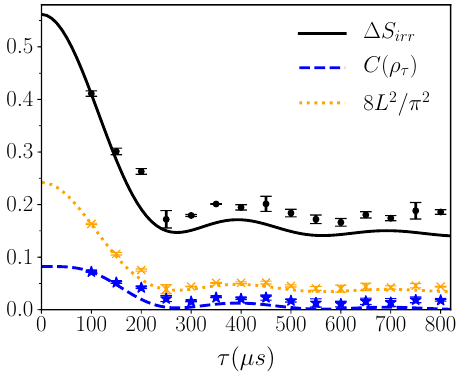}
\caption{(Color online) Same as Figure~\ref{3600_plot}, but with the energy gap
varied from 2000~Hz to 5000~Hz.}
\label{5000_plot}
\end{figure}
 
Experiments were performed by first initializing the spin temperature of both
the $^{13}C$ qubits to $(\beta_ih)^{-1}$=1580.2~Hz at thermal equilibrium.  In
order to probe system dynamics, it was driven out of equilibrium, on a time
scale varying from 100~$\mu$s to 800~$\mu$s.  During the driving process, the
initial energy gap was fixed at 2000~Hz and then changed to 3600~Hz; this
process was repeated with the final energy gap changed from 2000~Hz to 5000~Hz.
The final energy gap indicates how far the system is driven from
equilibrium; the larger the gap, the further the system moves away
from equilibrium.

The behavior of various quantities are plotted in Figure~\ref{3600_plot} and
Figure~\ref{5000_plot} for a final energy gap of 3600~Hz and 5000~Hz,
respectively.  
The irreversible entropy($\Delta S_{irr}$),
coherence(C($\rho(\tau)$)) and $8/\pi^2$ times the square of the Bures length
($8L^2/\pi^2$) are plotted in Figures~\ref{3600_plot} and \ref{5000_plot}, as a
function of the system driving time.  It is evident from the plots that for low
driving times, the system produces more entropy, since the driving time is far
away from the ideal quasi-static change in the system.  As the driving time is
increased, the irreversible entropy production gradually decreases.  since the
process is approaching a quasi-static process.  The coherence production
follows the same trend as the irreversible entropy production.  

The contribution of coherence to irreversible entropy production is more for a
final energy gap of 3600~Hz as compared to 5000~Hz.  This is evident from the
fact that the gap between the coherence curve and irreversible entropy curve is
less in the plot depicted in Figure~\ref{3600_plot} as compared to  the plot
depicted in Figure~\ref{5000_plot}.  The irreversible entropy produced is more
for a final energy gap of 5000~Hz (Figure~\ref{3600_plot}) as compared to a
final energy gap of 3600~Hz (Figure~\ref{5000_plot}), since the system is
driven further away from equilibrium for higher final energy gaps.  As the
driving time is increased, the contribution due to coherence is almost
negligible and the population difference between the states is the major factor
determining the irreversible entropy production.  Hence, we are able to
separate the irreversible entropy generated due to coherence production and the
mismatch in the population difference between the actual state and the
equilibrium state.  We have explored the validity of the inequality given in
Eqn.~\ref{general_inequality} in Figures~\ref{3600_plot} and \ref{5000_plot}
for a final energy gap of 3600~Hz and 5000~Hz, respectively.  It can be
observed that the inequality is always satisfied, which implies that the
irreversible entropy production is lower bounded.
Discrepancies between the theoretical curves and the experimental data can
be attributed to errors in pulse calibration, initial state preparation,
magnetic field inhomogeneities, and GRAPE pulse optimization.  These errors
lead to a final state being slightly different than the theoretical expected
state, which leads to the production of extra entropy.  Hence the experimental
data points are slightly above the theoretical curve (Figures~\ref{3600_plot}
and \ref{5000_plot}).
\section{Conclusions} 
\label{sec5} 
We have experimentally investigated an out-of-equilibrium quantum thermodynamic
process, where a system at thermal equilibrium is subjected to an external
driving force, which results in the production of entropy.  This irreversible
entropy has two separate contributions, one due to coherence generation after
external driving and the other due to a population mismatch between the actual
state and the equilibrium state.  We performed experiments by keeping the spin
temperature of the spins fixed and then driving the energy gap from an initial
2000~Hz to 3600~Hz and later upto a frequency of 5000~Hz.  
We found that coherence
generation follows the irreversible entropy trend, in that the amount of
entropy generated increases with increasing energy gap, and concomitantly, the
contribution to entropy generation due to coherence is decreased.  We also
verified the validity of a generalized Clausius inequality, which states that
the irreversible entropy production is lower-bounded by the Bures length
between the actual state and the equilibrium state. We were able to
time-separate the contribution of coherence to entropy generation and also
verified that the irreversible entropy generation is lower bounded for a
non-equilibrium process which is externally driven.

\begin{acknowledgments}
All experiments were performed on a Bruker Avance-III 600
MHz FT-NMR spectrometer at the NMR Research Facility at
IISER Mohali. 
K.S. acknowledges financial support from the 
Prime Minister's Research Fellowship (PMRF) scheme 
of the Government of India. 
\end{acknowledgments}


%

\end{document}